\pdfoutput=1

\documentclass[reprint,amsfonts,amsmath,amssymb,aps,pre,floatfix,10pt]{revtex4-1}

\usepackage{graphicx, amssymb, amsfonts, amsmath, epsf, color, enumerate, placeins, bm, floatrow}
\usepackage{overpic}

\usepackage{soul, color}
\usepackage[colorlinks, allcolors = blue]{hyperref}

\newcommand{\M}{\mathbf{M}}
\newcommand{\Q}{\mathbf{Q}}
\newcommand{\n}{\mathbf{n}}
\newcommand{\Fr}{\mathcal{F}}
\newcommand{\Rr}{\mathbb{R}}

\begin{document}

\title{Tailored Morphologies in 2D Ferronematic Wells}
\author{Konark Bisht$^{1}$}
  \author{Yiwei Wang$^{2}$}
  \email{ywang487@iit.edu}
  \author{Varsha Banerjee$^{1}$}
  \email{varsha@physics.iitd.ac.in}
  \author{Apala Majumdar$^{3}$}
  \email{apala.majumdar@strath.ac.uk}

  \affiliation{$^{1}$Department of Physics, Indian Institute of Technology, Delhi, Hauz Khas 110016, New Delhi, India. \\
    $^{2}$Department of Applied Mathematics, Illinois Institute of Technology, Chicago, IL 60616, USA. \\
  $^{3}$Department of Mathematics and Statistics, University of Strathclyde, G1 1XQ, UK.}


\begin{abstract}
  We focus on a dilute uniform suspension of magnetic nanoparticles in a nematic-filled micron-sized shallow well with tangent boundary conditions, as a paradigm system with two coupled order parameters. This system exhibits spontaneous magnetization without magnetic fields. We numerically obtain the stable nematic and associated magnetization morphologies, induced purely by the geometry, boundary conditions and the coupling between the magnetic nanoparticles and the host nematic medium. Our most striking observations pertain to domain walls in the magnetization profile whose location can be manipulated by the coupling and material properties, and stable interior and boundary nematic defects, whose location and multiplicity can be tailored by the coupling too. These novel morphologies are not accessible in uncoupled systems and can be used for new multistable systems with singularities and stable interfaces.
\end{abstract}

\maketitle

\section{Introduction} 
\label{sec:intro} 
Liquid crystals (LCs) are quintessential examples of soft materials or mesophases that are intermediate between conventional solids and liquids with a unique combination of long-range order and fluidity \cite{PG1995}. There are many different kinds of LCs, and we focus on nematic liquid crystals (NLCs) with long-range orientational order that manifests in nematic ``directors'' or locally distinguished directions of averaged molecular alignment \cite{PG1995, stewart}. The directional nature of NLCs makes them susceptible to external fields, incident light, temperature and foreign inclusions, e.g., colloids. 
Historically, most NLC-based applications have relied on their dielectric anisotropy, i.e., direction-dependent response to electric fields, e.g., the thriving liquid crystal display industry. The anisotropy in the NLC magnetic susceptibility is typically much smaller ($\sim10^{-6}$), in some cases $7$ orders of magnitude smaller than the dielectric anisotropy \cite{PG1995, AM2013, AM2017} and consequently, the exciting field of magnetic phenomena in NLCs and partially ordered materials remains relatively open.
    
In the 1970s, Brochard and de Gennes suggested that the addition of magnetic nanoparticles (MNPs), i.e., nanoparticles with magnetic moments, to a nematic medium can generate a spontaneous magnetization without any external fields and such systems were referred to as ferronematics in their pioneering work \cite{FB1970}. The magnetic moments of MNPs are influenced by the ambient nematic directors (and vice-versa) due to the surface-induced coupling between them. Magnetic moments preferentially align either parallel or perpendicular to the nematic director depending on the surface treatment of nanoparticles. The suspension shows spatial macroscopic magnetization in the absence of fields. 
This spontaneous polar magnetization can, in turn, substantially enhance magneto-optic responses in NLC systems. In 2013, Mertelj {\it et al.} designed the first such stable ferronematic suspension using barium hexaferrite (BaHF) magnetic nanoplatelets in pentylcyano-biphenyl (5CB) LCs \cite{AM2013}. There have been several challenges related to MNP aggregation and flocculation, but the platelet shape, high magnetocrystalline anisotropy and polydispersity of the MNPs and the MNP-NLC interactions (dipolar/quadrupolar) can be exploited to stabilize such ferronematic suspensions \cite{AM2013, AM2017}. Since then, there has been a wave of interest in the physical, optical and rheological properties of ferronematics \cite{AM2013, AM2017, AH2015, QL2016, PA2017}, in designing new composite materials \cite{PA2017},  biaxial ferronematics \cite{QL2016}, chiral ferronematics \cite{PA2017}, easily switchable MNP-NLC systems with small magnetic fields \cite{QL2016} and even creating topological solitons in ferronematics \cite{PA2017}. From a purely scientific point of view, the coupled MNP-NLC systems give us access to new singular structures, new theoretical frameworks and exotic morphologies which are inaccessible in uncoupled systems. From an applications point of view, leading experimentalists propose that these MNP-NLC systems have new magneto-mechanical and magneto-optic effects with potential applications in photonics \cite{PA2017, MM2017}, display devices \cite{CC2016},  optics \cite{PR2015, AM2014, AH2015}, telecommunications \cite{AM2017},  microfluidics \cite{ AM2017} and smart fluids \cite{RS2015, TP2017}.  

In \cite{KB2019}, we study a model problem of a ferronematic suspension in a nematic-filled channel in a one-dimensional setting. There are two variables - a nematic order parameter and a magnetic order parameter, induced by the suspended nanoparticles, without any external fields, with Dirichlet conditions for both order parameters on the bounding surfaces. We work in a continuum framework and use the one-dimensional geometry, choice of boundary conditions and the nemato-magnetic coupling to give examples of tailored inhomogeneous morphologies and domain wall formation in such systems, also looking at the effects of temperature in destabilizing domain walls. However, the one-dimensional nature of the problem naturally limits the solution landscape for both the nematic and magnetic order parameters. In this paper, we study a two-dimensional benchmark and highly informative problem with a much richer solution landscape - a dilute uniform suspension of MNPs in an NLC-filled square well, motivated by the experimental work in \cite{CT2007}. We assume that the well surfaces are treated to induce planar degenerate or tangent boundary conditions, i.e., the nematic molecules in contact with these surfaces are in the plane of the surfaces. Following previous work on this problem \cite{CT2007, MR2017, YW2019}, we impose Dirichlet conditions on the lateral surfaces for the nematic molecules and strong anchoring on the top and bottom surfaces, that enforce planar degenerate conditions on these surfaces without any preferred directions. As in \cite{DG2017}, one can rigorously prove that in the thin-film limit (when the well height is much smaller than the cross-sectional dimensions), we can study planar two-dimensional nematic profiles on the square cross-section, that are invariant across the height of the well. We adopt this reduced approach and study coupled ferronematic systems on a square domain with Dirichlet or fixed tangent conditions for both the nematic molecules and the spontaneous magnetization on the square edges (also see \cite{AM2013}). Whilst Dirichlet tangent conditions are well accepted in the nematic framework \cite{CT2007, CL2012, JW2018}, the correct choice of boundary conditions for the spontaneous magnetization remains open. In \cite{MS2016}, the authors argue that tangent boundary conditions naturally arise for spontaneous magnetization from energetic considerations. More generally, Dirichlet conditions for the spontaneous magnetization can be achieved by experimentally controllable ferromagnetic walls or superparamagnetic walls \cite{private_communication_igor_musevic} 
or by applying an external field to fix the position and orientation of the MNPs on the edges, and then removing the field disallowing further reorientation of the MNPs on the edges. Given that ferronematics are relatively nascent, a theoretical study of nemato-magnetic coupling with different types of boundary conditions (such as the model problem studied here) opens new avenues for experimental investigations. This square system has received significant attention in the purely nematic case (see for example \cite{MR2017}), where up to $21$ different nematic states have been reported. It is natural to ask how this well-studied system responds to the inclusion of MNPs? In particular, the nematic-filled square well is known to be experimentally bistable \cite{CT2007} without any external electric fields. Natural questions are - is the coupled MNP-NLC system multistable too, can we stabilize interior nematic defects with the MNP-NLC coupling, can this MNP-NLC system be an example of a liquid crystal device controlled by magnetic fields as opposed to electric fields?  We partially address some of these questions in this paper and in doing so, unravel several complex phenomena in a relatively simple ferronematic set-up. These questions are not specific to ferronematics, but also apply to a coupled system with competing nematic and polar order in the absence of external fields.

For a dilute ferronematic suspension as considered here, there are two continuous macroscopic order parameters - \cite{AM2013,  GZ2018, MC2014}: (i) the NLC $\Q$-tensor order parameter contains information about the orientational anisotropy of the NLC; and (ii) the magnetization vector, $\M$, which is the spatially averaged magnetic moment of the suspended MNPs at every point inside the domain. We do not account for any dipolar/quadrupolar interactions between the magnetic nanoparticles (also see \cite{AM2016}) explicitly since experimentalists report that the attractive dipolar forces between magnetic moments are counteracted by the repulsive nematic-mediated quadrupolar interactions between the nanoparticles, for stable suspensions. Further we are working in a continuum framework, and the nematic-MNP and MNP-MNP interactions are absorbed by a coupling energy \cite{AM2013, AM2017}. 

More precisely, we build on the phenomenological approach in \cite {AM2016} and model the stable $\left(\Q, \M\right)$ profiles as minimizers of an appropriately defined free energy, that contains a magneto-nematic coupling energy.  Mathematically, this is equivalent to solving a coupled system of nonlinear partial differential equations with several technical difficulties. There are four key phenomenological parameters that contain information about intrinsic length scales, material properties and the magneto-nematic coupling. In fact, by tuning these parameters, we numerically observe stable domain walls in $\M$(analogous to experimental results in \cite{AM2013, QL2016, MS2016}), displace the domain walls in $\M$, generate exotic splay-twist $\M$ profiles and crucially, stabilize interior nematic point defects without external fields. In fact, the tuning parameters allow us to control the locations and multiplicities of the nematic defects. In particular, if we can experimentally realize stable defect structures in coupled MNP-NLC systems, these may provide new routes to probe universal defect structures in different branches of physics and create self-assembled patterns of nematic defects and magnetic domain walls without any external fields, and our results are a forward step in that direction. In Section~\ref{sec:theory}, we review the theoretical framework and governing partial differential equations; in Section~\ref{sec:results}, we present our numerical results and in Section~\ref{sec:conclusions}, we state the main conclusions and directions for future work.
        
\section{Theory}
\label{sec:theory}

Our domain is a square with edge length $L$ i.e.
$$ \Omega = \left\{ \left(x, y \right) \in \Rr^2; 0 \leq x,y \leq L \right\} $$
where $L$ is typically on the scale of microns. 
For this 2D problem, we assume that the MNPs primarly lie in a plane with favoured in-plane alignment of magnetic moments \cite{AM1994}. Therefore, we take the magnetization vector, $\M$, to be a 2D vector and in particular, $\M$ can have variable magnitude, including $\M =0$. In particular, domains walls in $\M$ belong to the zero set of $\M$ or more generally, regions of small $|\M|$. In our reduced approach, the $\Q$-tensor is a symmetric, traceless $2\times 2$ matrix whose leading eigenvector $\n$ (with the largest eigenvalue which is necessarily positive)  models the locally preferred direction of NLC alignment at every point in space.  We refer to $\n$ as the 2D nematic director i.e.  
$\n = \cos\phi\hat{x}+  \sin\phi\hat{y}$, where $\phi$ is the angle between $\n$ and $\hat{x}$, $\hat{x}$ and $\hat{y}$ are the coordinate unit vectors in the plane. Then $\Q$ can be written as $\Q =  S\left[2\n\otimes\n- {\bf I} \right]$, and the scalar order parameter $S$ is a measure of the degree of orientational order and fluctuations about $\n$ \cite{CL2012, MR2017}; a nematic defect corresponds to regions of low order with $S \approx 0$. Since $\Q$ is traceless,  one can write
$$\Q = \begin{pmatrix} Q_{11} &  Q_{12} \\ Q_{12} & - Q_{11} \\ \end{pmatrix}$$
and easily verify that $\mbox{ Tr}\Q^2 = {\vert\Q
  \vert}^2= 2(Q_{11}^2+Q_{12}^2) = 2S^2$ and  $\mbox{ Tr} \Q^3 = 0$. The generalized free energy density for this composite system has three contributions \cite{PG1995, HP2001, KB2019, AM2016}:

\begin{equation}\label{eq:LGLDG}
  \begin{aligned}
    \Fr &= \frac{K}{2}\sum_{ij}{\vert\nabla Q_{ij}\vert}^2+\frac{A}{2} \mbox{Tr} \Q^2+\frac{C}{4} \left(\mbox{Tr}\Q^2\right)^2 \\
    &+\frac{\kappa}{2}\sum_{i}{\vert\nabla M_i\vert}^2+\frac{\alpha}{2}{\vert\M\vert}^2+\frac{\beta}{4}{\vert\M\vert}^4 \\
    &-\frac{\gamma\mu_0}{2}\sum_{ij}Q_{ij}M_iM_j.
  \end{aligned}
\end{equation}

The first line is the NLC Landau-de Gennes free energy density, the next line is the Ginzburg-Landau free energy density for the magnetization and the last line is the magneto-nematic coupling energy density \cite{SB1995MSEC, HP2001}. The Landau coefficient $A = \bar{A}\left(T-T^*\right)$, where $\bar{A}$ is a positive constant and $T^*$ is a characteristic transition temperature for NLC. Similary, $\alpha= \bar{\alpha}\left(T-T_{c}^M\right)$, where $\bar{\alpha}$ is a positive constant and $T_{c}^M$ is a critical temperature for the spontaneous magnetization. The parameters $C$ and $\beta$ are positive material-dependent constants whereas $K$ and $\kappa$ are the elastic constants, related to NLC elasticity and magnetic stiffness respectively. The elastic energy density for $\M$ is included in general continuum energies for ferronematics as in \cite{AM2017} and can be viewed as regularization term that penalizes short-range variations in $\M$. This term prevents arbitrary rotations in {\bf{M}} without an energetic penalty. Lastly, $\gamma$ is a MNP-NLC coupling parameter \cite{SB1995MSEC} such that positive values of $\gamma$ coerce $\n$ and $\M$ to be parallel to each other whereas negative values of $\gamma$ coerce $\n$ and $\M$ to be perpendicular to each other. This can be roughly seen by the following calculation,

\begin{equation}
  -\frac{\gamma\mu_0}{2}\sum_{ij}Q_{ij}M_iM_j =\gamma\mu_0S{\vert\M\vert}^2\left( \frac{1}{2} - \cos^2\theta \right),
  \label{coupling}
\end{equation}
where $\theta$ is the angle between the nematic director $\n$ and the magnetization vector $\M$; see \cite{HP2001} for more details.  We consider the simplest form of the coupling term that is adequate to stabilize bulk ferronematic phases \cite{HP2001}. For $S$ constant, this effectively reduces to $\left(\n\cdot\M \right)^2$ as used in \cite{AM2013}. While higher order magneto-nematic coupling terms are required for the study of critical behaviour or phase diagrams, the simple cubic term suffices to capture stable ferronematic states as studied here.

These phenomenological parameters are typically estimated from experimentally measured quantities \cite{PS1975}. The Landau coefficients $A$, $B$ and $L$ are related to experimentally measured quantities like the isotropic-nematic transition temperature, the latent heat of transition and the order parameter \cite{EP2012}. Similarly, the coefficients $\alpha$, $\beta$ and $\kappa$ can be evaluated from the measurements of magnetization and susceptibility \cite{PH2015}. The coupling constant, $\gamma$, has been estimated from the reversal fields of hysteresis loops in \cite{AM2013}. In principle, it is possible to estimate the parameters above, but the available experimental data on ferronematics is extracted in the presence of external magnetic fields. As the latter influences response functions and other characteristics, we cannot obtain reliable estimates of the phenomenological parameters above with zero magnetic fields at this juncture. However, we hope that our work will motivate new investigations on these lines for ferronematics.
    
We rescale Eq.~(\ref{eq:LGLDG}) by defining $[Q^\prime_{11},Q^\prime_{12}] = \sqrt{2C/\vert A\vert}[Q_{11}, Q_{12}]$, $[M^\prime_1$, $ M^\prime_2] = \sqrt{\beta/\vert\alpha\vert}$ $[M_1$, $M_2]$, $[x^\prime$, $y^\prime]= $ $[x$, $y]/L$ (where $L$ is the length of the side of the well), and $\Fr^\prime = \Fr C/A^2$. 
The Euler-Lagrange (EL) equations associated with the dimensionless free energy density are given by:
\begin{equation}\label{EL}
  \begin{aligned} 
    &\ell_1 \nabla^2 Q_{11} - \widetilde{Q}Q_{11} +\frac{c}{2}\left(M_1^2-M_2^2\right) = 0,\\
    &\ell_1\nabla^2 Q_{12} -\widetilde{Q}Q_{12} +  cM_1M_2 = 0,\\
    & \xi\left(\ell_2\nabla^2M_1 -  \widetilde{M}M_1\right)+ c\left(Q_{11}M_1+Q_{12}M_2\right) = 0, \\
    &\xi\left(\ell_2\nabla^2M_2  - \widetilde{M}M_2\right) +  c\left(Q_{12}M_1-Q_{11}M_2\right) = 0, \\
  \end{aligned}
\end{equation}
where  $\widetilde{Q}=\left(\mbox{ Tr}{\Q}^2/2- 1\right)$, $\widetilde{M} =  \left({\vert\M\vert}^2-1\right)$ and
\begin{equation*}
  \ell_1 = \frac{K}{\vert A\vert L^2}; \ell_2 = \frac{\kappa}{\vert\alpha\vert L^2};\xi = \frac{C}{\vert A\vert^2}\frac{{\vert\alpha\vert}^2}{\beta};c = \frac{\gamma\mu_0}{\vert A\vert}\sqrt{\frac{C}{2\vert A\vert}}\frac{\vert\alpha\vert}{\beta}.
\end{equation*}

\begin{figure}[!tb]
  \centering
  \includegraphics[width = \linewidth]{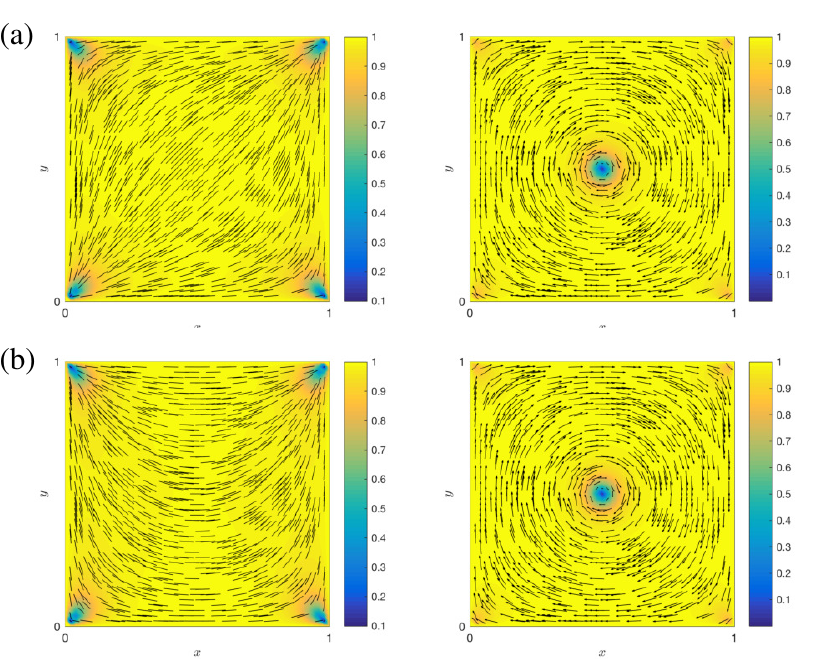}
  \caption{Nematic [left] and  magnetic configurations [right]  for  $\ell_1 = \ell_2 = 0.001$, $c = 0$: (a) $D_N$, (b) $R_N$. The color bars denote the values of $S$ and $|{\bf M}|$ respectively.}\label{Uncoupled}
\end{figure}
    
There are four dimensionless phenomenological constants above. We work with low temperatures so that $A, \alpha <0$. The parameter $c$ is the coupling constant; the sign of $c$ has the same meaning as the sign of $\gamma$ above. 
The parameters $\sqrt{\ell_1}$ and $\sqrt{\ell_2}$ set the scale of curvature for $\Q$ and $\M$ respectively; for simplicity we set them to be identically equal to $\sqrt{\ell}$, which is a physically relevant choice since $\sqrt{\ell_1}$ and $\sqrt{\ell_2}$ are defined as ratios of parameters. In our simulations, unless otherwise stated, we take $\ell$ = 0.001, which is a benchmark value motivated by previous studies for uncoupled systems with $c=0$ in \cite{CL2012}. The fourth parameter, $\xi$ is a measure of the strength of the magnetic energy relative to the nematic energy, i.e., larger values of $\xi$ will coerce the composite system to minimize the magnetic energy in Eq.~(\ref{eq:LGLDG}) so that the magnetization profile $\M$ will strongly tailor the $\Q$ - profile but not necessarily the other way around (i.e. $\M \to \Q$) (at least for minimizers of (\ref{eq:LGLDG})). Similarly, for very small values of $\xi$, minimizers of the composite system are less influenced by the magnetic energy density in Eq.~(\ref{eq:LGLDG}) and in this limit, the $\M$ profiles are tailored by the $\Q$ profiles and not strongly in the other direction (i.e. $\Q\to\M$). Both limits, $\xi\to 0$ and $\xi\to\infty$, physically describe the one-way coupling of $\Q\to\M$ and $\M\to\Q$ respectively.

We work with Dirichlet boundary conditions for $\Q$ and $\M$ respectively on the re-scaled square edges,  $x =0,1$ and $y = 0,1$ and our choices are guided by earlier experimental and theoretical works, which assume that $\n$ is constrained to be tangent to the square edges \cite{CT2007, YY2007, YY2009, CL2012, SK2014, AL2014}. 
The tangent conditions require that $Q_{11} = -1$,  $Q_{12} = 0$ at $x = 0,1$  and $Q_{11} = 1$,  $Q_{12} = 0$ at $y=0,1$ (this is equivalent to fixing $\n = \left(\pm 1, 0 \right)$ on the edges $y=0,1$ and $\n = \left(0, \pm 1 \right)$ on the edges $x=0,1$). We assume that $S=1$ on the edges, by analogy with previous work on uncoupled systems. Since we have no experimental data for the boundary values of $S$ in the context of ferronematics, we believe that $S=1$ is a good starting point for theoretical studies on these lines. 
For $\M$, we assume $\M=(0,1)$ at $x=0$; $\M=(0,-1)$ at $x=1$; $\M=(-1,0)$ at $y=0$; $\M=(1,0)$ at $y=1$. These are topologically non-trivial tangent Dirichlet conditions by construction, motivated by similar experimental observations in confined ferronematic systems by Shuai et al. \cite{MS2015} that require $\M$ to rotate by $2\pi$ radians along the boundary, so that the $\M$ profile necessarily has an interior defect with $\M = 0$. This is a plausible choice in the strong anchoring limit with positive $\gamma$ (see \cite{HP2001, MS2016}), since $\M$ is either parallel or anti-parallel to $\n$ on the square edges, and we speculate that there may be experimental methods to fix $\M$ on the edges, even with negative values of $\gamma$. In fact, the topologically non-trivial boundary conditions for $\M$ stabilize interior nematic defects in our numerical results, in certain parameter regimes. We could also impose natural boundary conditions for ${\bf M}$ which are experimentally realisable and essentially imply that we do not prescribe any boundary conditions for ${\bf M}$. Preliminary numerical investigations show that in the case of natural boundary conditions for ${\bf M}$, the ${\bf M}$-profile is tailored by the ${\bf Q}$ profile for positive and negative $c$ and we may not achieve two-way coupling between ${\bf Q}$ and ${\bf M}$, as can be achieved by our choice of Dirichlet conditions for ${\bf M}$.

\begin{figure}[!tb]
  \centering
  \includegraphics[width = \linewidth]{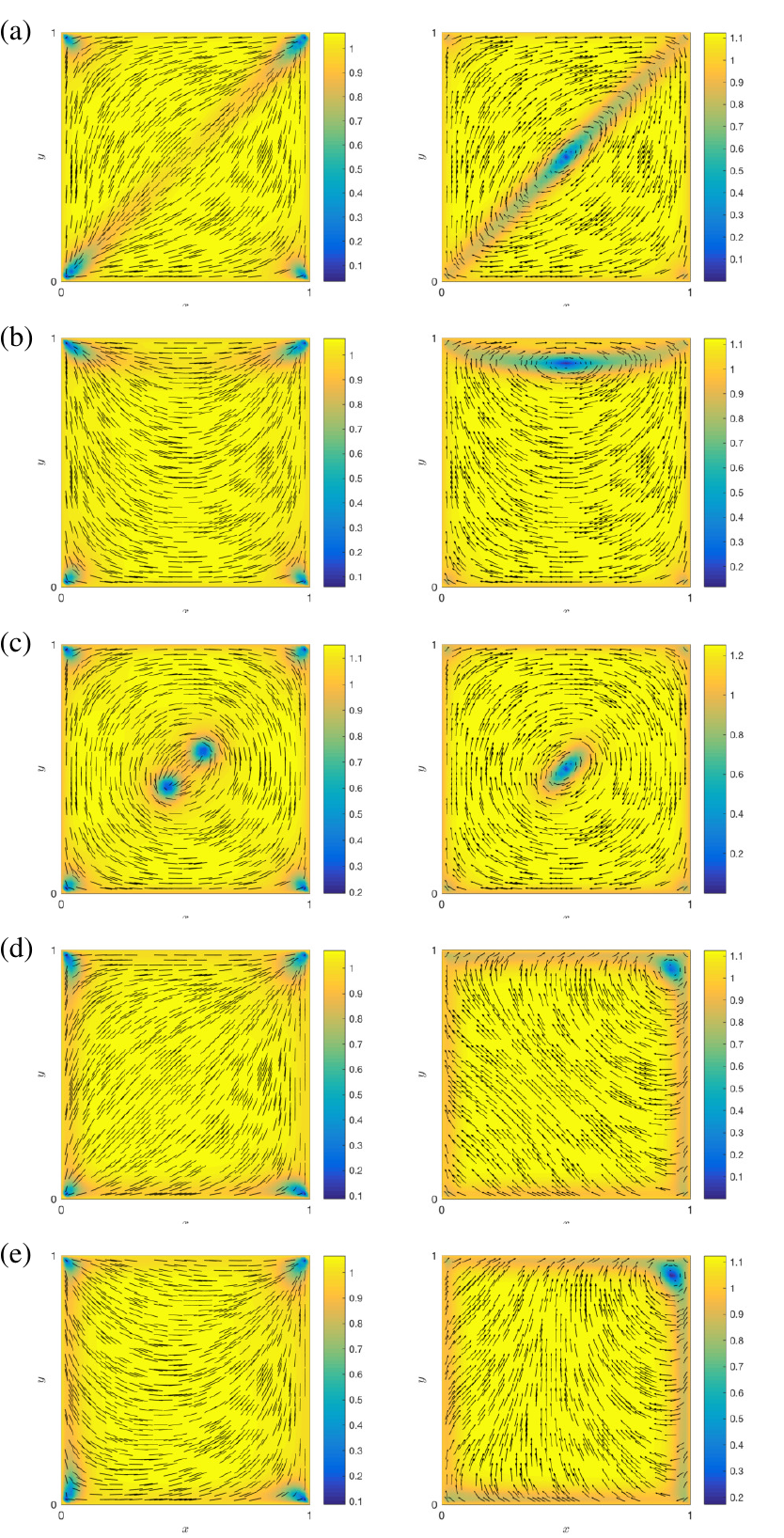}
  \caption{Nematic and  magnetic configurations  for  $\ell_1 = \ell_2 = 0.001$,  $\xi = 1$ and (a) $c = 0.25$, $\left( \Q_{D, 0.25}^{1}, \M_{D, 0.25}^{1} \right)$; (b) $c = 0.25$, $\left( \Q_{R, 0.25}^{1}, \M_{R, 0.25}^{1} \right)$; (c) $c = 0.5$, $\left( \Q_{D^{*}, 0.5}^{1}, \M_{D^{*}, 0.5}^{1} \right)$; (d) $c = -0.25$, $\left( \Q_{D, -0.25}^{1}, \M_{D, -0.25}^{1} \right)$; (e) $c = -0.25$, $\left( \Q_{R, -0.25}^{1}, \M_{R, -0.25}^{1} \right)$. }\label{fig:xi_1}  
\end{figure}

\begin{figure*}[!tb]
  \centering
  \includegraphics[width = \linewidth]{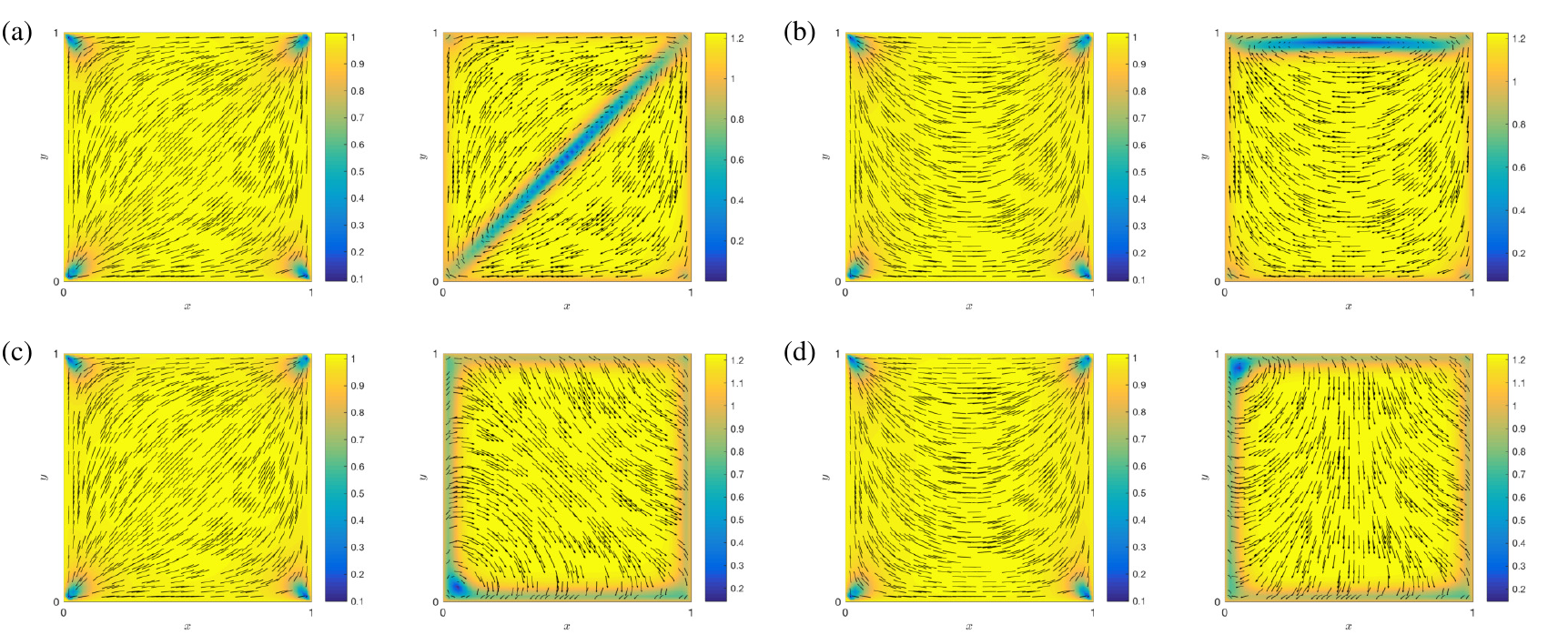}
  \caption{Nematic and  magnetic configurations  for  $\ell_1 = \ell_2 = 0.001$, $\xi = 0.1$ and (a) $c = 0.05$, $\left( \Q_{D, 0.05}^{0.1}, \M_{D, 0.05}^{0.1} \right)$; (b) $c = 0.05$, $\left( \Q_{R, 0.05}^{0.1}, \M_{R, 0.05}^{0.1} \right)$; (c) $c = -0.05$, $\left( \Q_{D, -0.05}^{0.1}, \M_{D, -0.05}^{0.1} \right)$; (d) $c = -0.05$, $\left( \Q_{R, -0.05}^{0.1}, \M_{R, -0.05}^{0.1} \right)$.}\label{fig:xi_0.1}
\end{figure*}

    
The critical points of this system are solutions of EL equations (\ref{EL}), which are computed numerically by standard finite-difference method (five-point formula for $\Delta$) and Newton's Method \cite{iserles2009first, YW2019}. In order to track the complex solution landscape, we apply the deflation technique for nonlinear problems \cite{PF2015, MR2017, YW2019}. The solution stability can be checked by looking at the smallest eigenvalue $\lambda_1$ of the Hessian matrix of the discretized free energy \cite{MR2017, GC2019}. A solution is locally stable if $\lambda_1 > 0$. In what follows, we first discuss the uncoupled system with $c = 0$, then discuss the effects of the MNP-NLC coupling with different values of $\xi$ ($ = 0.1, 1$ and $10$), with both $c>0$ and $c<0$. Some of our most interesting results pertain to how we can tune the system properties with $c$ and $\xi$, opening new vistas of scientific and experimental possibilities.

\section{Numerical Results}
\label{sec:results}

NLC-filled square wells (corresponding to $c=0$) are well studied; see \cite{CT2007, CL2012, SK2014, HK2015, AM2016, MR2017, JW2018, YW2018, XY2018, GC2019}. There are two stable nematic equilibria: $D_N$ and $R_N$ \cite{CT2007, CL2012}. The solution $D_N$ has a diagonally aligned nematic director whereas the nematic director rotates by $\pi$ radians between a pair of opposite edges for the $R_N$ state. We plot the $D_N$ and $R_N$ states in Fig.~\ref{Uncoupled}, along with the scalar order parameter (read by the colour chart). Both the $D_N$ and $R_N$ states have zero winding number, i.e. the corresponding $\n$ has no net rotation around the square boundary. We also plot the $\M$ profile with $c=0$. As expected, we see a distinct $+1$-degree vortex at the square centre (with $\M = {\bf 0}$) consistent with the topologically non-trivial boundary conditions for $\M$. We point out that $\M$ has a direction whereas the $\n$-field is a director field without a direction. We plot $\vert\M\vert$ with the colour bar to highlight the reduction in $\vert\M\vert$ at the vortex centre. 
    
\begin{figure*}[t]
  \centering
  \includegraphics[width = \linewidth]{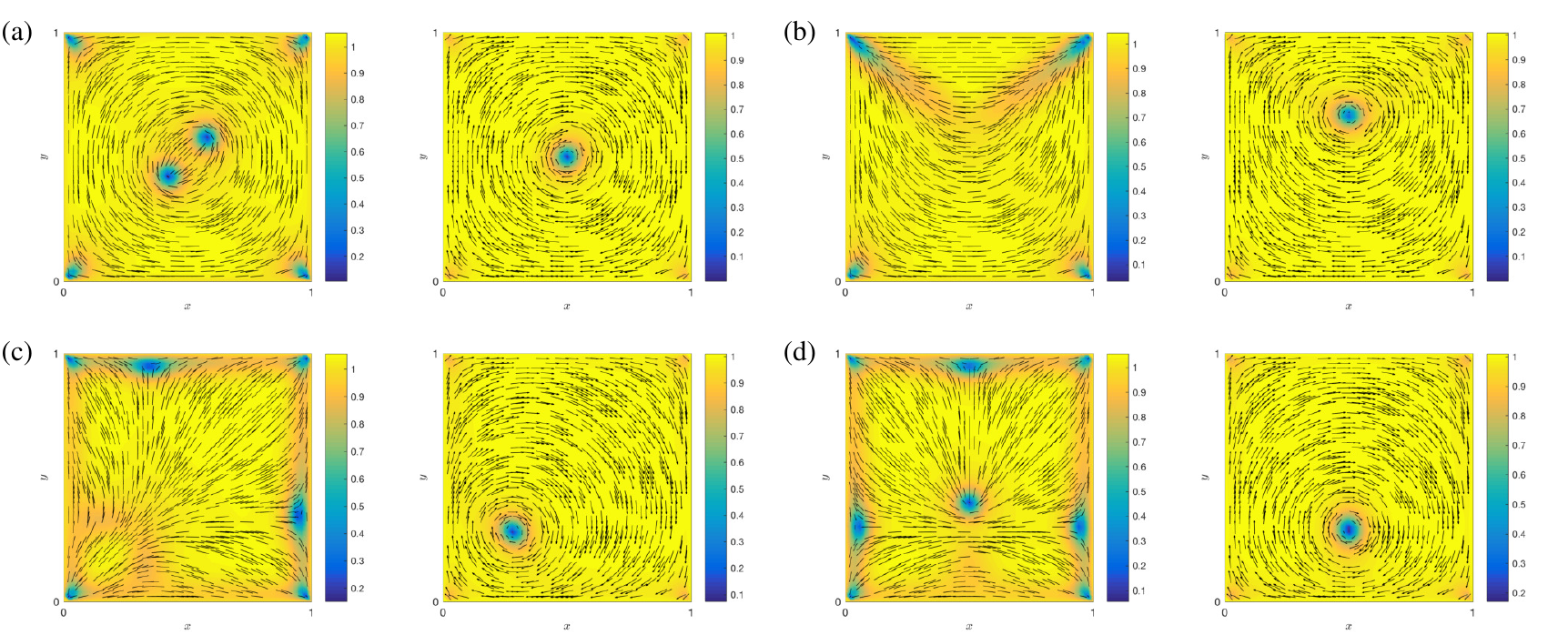}
  \caption{Nematic and  magnetic configurations  for  $\ell_1 = \ell_2 = 0.001$, $\xi = 10$ and (a) $c = 0.25$, $\left( \Q_{D, 0.25}^{10}, \M_{D, 0.25}^{10} \right)$; (b) $c = 0.2$, $\left( \Q_{R, 0.2}^{10}, \M_{R, 0.2}^{10} \right)$; (c) $c = -0.25$, $\left( \Q_{D, -0.25}^{10}, \M_{D, -0.25}^{10} \right)$; (d) $c = -0.25$, $\left( \Q_{R, -0.25}^{10}, \M_{R, -0.25}^{10} \right)$.}\label{fig:xi_10}
\end{figure*}
    
Next, we consider coupled systems ($c \neq 0$) with $\xi =1$, where the nematic and magnetic energy densities are of comparable importance for energy minimizers or locally stable solutions $(\Q, \M)$ of Eq.~(\ref{EL}). In Fig.~\ref{fig:xi_1}(a)-(b), we plot two numerically computed stable solutions of Eq.~(\ref{EL}) with $\xi = 1$ and $c=0.25$, denoted by $(\Q_{D,0.25}^{1}, \M_{D,0.25}^1)$ and $(\Q_{R,0.25}^1, \M_{R,0.25}^1)$, respectively. Here, the superscripts specify the value of $\xi$, and the subscripts $D$ and $R$ indicate the numerical solutions computed with $D_N$ and $R_N$ as initial numerical guesses for the $\Q$-solution respectively, and the numerical value in the subscript is the value of $c$. 
In this regime, the $\Q$ profile does not develop any interior defects (with $S\approx 0$) since $\xi$ is unity. Rather, we see that $ \M_{D,0.25}^1$ exhibits a smeared out vortex along the square diagonal, to induce co-alignment between $\n$ and $\M$, creating an interior domain wall with $\M \approx 0$. Similar remarks apply to the pair $\left(\Q_{R, 0.25}^1, \M_{R,0.25}^1 \right)$ where we observe a distinct domain wall (with  $\vert\M\vert\approx 0$) containing a smeared out vortex near one of the square edges, that induces a rotated-like, $\M_R$, profile away from the domain wall. There is a corresponding reduction in $S$ for the $\Q_R$ profile too, along this square edge, tailored by the domain wall in $\M_{R,0.25}^1$. Interestingly, there is some evidence of two-way coupling for $\xi=1$ when $c$ is large enough. An example for $c = 0.5$, labelled by $(\Q_{D^*,0.5}^{1}, \M_{D^*,0.5}^1)$, is shown in Fig. \ref{fig:xi_1}(c). Here, the subscript $D^{*}$ indicates that this is another computable solution using $D_N$ as the initial guess for the $\Q$-profile and the deflation technique.  There are emergent point defects of charge $+1/2$ in the $\Q$-solution tailored by the interior vortex in ${\bf M}_{D^{*},0.5}^1$. We also consider negative coupling for $\xi = 1$ and compute two locally stable critical points for $c = -0.25$, denoted by $(\Q_{D,-0.25}^{1}, \M_{D,-0.25}^1)$ and $(\Q_{R,-0.25}^1, \M_{R,-0.25}^1)$ respectively, see Fig. \ref{fig:xi_1}(d)-(e). In both cases, $\n$ retains the diagonal and rotated profiles respectively, and $\M$ distorts to be perpendicular to the corresponding $\n$. Notably, the vortices in $\M$ migrate to a square vertex, and this may have interesting optical consequences for experiments.  
    
In Fig.~\ref{fig:xi_0.1} (a)-(d), we set $\xi = 0.1$ and look at $c=0.05$ and $c= -0.05$ respectively. Again there are multiple critical points but we only illustrate stable pairs $\left( \Q_{D, \pm 0.05}^{0.1}, \M_{D,\pm 0.05}^{0.1} \right)$ and  $\left( \Q_{R, \pm 0.05}^{0.1}, \M_{R,\pm 0.05}^{0.1} \right)$. Qualitatively, the profiles look similar to stable profiles with $\xi = 1$ with little coupling effect on $\n$ (since $\xi$ is quite small). For positive $c$, as in Fig.~(\ref{fig:xi_0.1})(a)-(b), we see a distinct domain wall (with  $\vert\M\vert\approx 0$) along one of the square diagonals in $\M_{D, 0.05}^{0.1}$ or near one of the square edges in $\M_{R, 0.05}^{0.1}$. This gives us excellent control on magnetic domain walls, their structure, location and stability. Equally, for $c = -0.05$, shown in Fig.~(\ref{fig:xi_0.1})(c)-(d), $\M_{D, -0.05}^{0.1}$ and $\M_{R, -0.05}^{0.1}$, reorient to be perpendicular to the corresponding nematic directors, the vortices migrate to one of the square vertices and interestingly, we see partial domain walls in $\M$ along pairs of adjacent square edges. In this regime, $\M$ is more susceptible to $\Q$ and we can use $\Q$ to tailor $\M$ effectively.
    
Finally, in Fig.~\ref{fig:xi_10} (a)-(d), we set $\xi = 10$ and study the effects of both positive and negative coupling, through the solution pairs $\left(\Q_{D, 0.25}^{10}, \M_{D, 0.25}^{10} \right)$, $\left(\Q_{R, 0.2}^{10}, \M_{R, 0.2}^{10} \right)$,
$\left(\Q_{D, -0.25}^{10}, \M_{D, -0.25}^{10} \right)$ and $\left(\Q_{R, -0.25}^{10}, \M_{R, -0.25}^{10} \right)$ respectively.
For $\xi = 10$, the $\M$ profiles strongly tailor the corresponding $\n$ profiles as expected i.e., the corresponding $\M$ vector fields retain the interior vortex in all four cases in Fig.~\ref{fig:xi_10} (a) - (d). For $\xi = 10$ and $c=0.25$ [Fig.~\ref{fig:xi_10} (a)], the nematic director $\n_{D, 0.25}^{10}$ loses the diagonal profile and exhibits two distinct $+1/2$ interior point defects, following the $\M_{D, 0.25}^{10}$ profile.  For $c = 0.2$, shown in Fig.~\ref{fig:xi_10} (b), we observe a clear displacement of the interior vortex in $\M_{R, 0.2}^{10}$, in response to $\n_{R, 0.2}^{10}$. This displaced vortex state is not stable for stronger coupling, say $c = 0.25$.
    
 For $c=-0.25$, two defects appear near the square edges in $\n_{D, -0.25}^{10}$ and the interior vortex in ${\bf M}$ migrates towards the left-down corner in  $\left( \M_{D, -0.25}^{10} \right)$ [See Fig.~\ref{fig:xi_10} (c)], and  in $\left(\Q_{R, -0.25}^{10}, \M_{R, -0.25}^{10} \right)$, the interior $\M$-vortex migrates vertically downwards in $\M_{R, -0.25}^{10}$ whilst we observe three defects on the square edges and one interior $+1/2$ defect in the corresponding $\n_{R, -0.25}^{10}$ profile. Whilst we cannot give detailed explanations about the appearance and multiplicity of these nematic defects, they are stable and arise naturally from energetic and topological considerations without external fields. 
 
 \section{Conclusions}
 \label{sec:conclusions}
    
    We report new and exotic morphologies in an MNP-NLC square system, that exhibit defects in both $\n$ and $\M$, with rich spatial inhomogeneities. These textures are stabilized by an interplay between the coupling parameter, $c$, and material and temperature-dependent parameter, $\xi$. The coupling parameter $c$ is largely a material property and in some cases, acts as an external magnetic field, i.e., we can displace magnetic domain walls by varying $c$. The parameter, $\xi$, has not been highlighted in the literature due to limited theoretical studies. The material-dependent constants, $C$ and $\beta$ and the temperature-dependent parameters, $A$ and $\alpha$, can be tuned to control $\xi$. We focus on static ferronematic equilibria without any external magnetic fields which address the pivotal question - what are the observable states in this coupled system, with non-polar nematic and polar magnetic order? The most likely state is the global energy minimizer, and we defer energy comparisons to future work. 
    
    Our work in a simple 2D setting actually captures (to some extent) the experimentally reported complex magnetic domain walls, nematic twist walls and other defects in a three-dimensional ferronematic-filled rectangular capillary in \cite{MS2016}. Further, we can adapt our theoretical methods to study defect lattices in ferronematics (also see reported in \cite{AH2015, QL2016}) and defects as binding sites for new materials design. The dynamic counterparts of our static study are equally rewarding, e.g., persistent vortices (which are essential for nano and microscale mixing applications) in microfluidic channels or even applications in electrokinetics \cite{CC2018}, and in tailored colloidal assemblies \cite{IM2006, MR2009, AN2013, YW2017, YW2018, SMondal2018}. Last but not least, we plan to study the interaction of these simple systems with external magnetic fields. With an external magnetic field, both $\n$ and $\M$ will couple to each other and with the external field and this coupling can either enhance or suppress the effects of $c$ and $\xi$, or lead to completely different morphologies. For example, magnetic domain walls may have different laws of motion with magnetic fields or one may even see stable nematic line defects in the interior, tailored by the external field. Equally, we may be able to switch between the reported ferronematic equilibria by applying relatively small magnetic fields. If the external field is non-planar, a three-dimensional approach is needed for both $\Q$ and $\M$. This would be a new ball game with fundamentally new scientific implications.
    
    \acknowledgments
    KB acknowledges CSIR (India), for financial support under Grant No. 09/086(1208)/2015-EMR-I. The authors gratefully acknowledge partial financial support from DST-UKIERI and the HPC facility of IIT Delhi for the computational resources. YW would also like to thank the Department of Applied Mathematics at Illinois Institute of Technology for their generous support and a stimulating environment. We thank Professor Igor Mu{\v s}evi{\v c} for illuminating discussions about relevant choices of the boundary conditions. AM and the team started working on this project when AM was a faculty member at the University of Bath. AM and her co-authors thank the University of Bath for a stimulating research environment and their hospitality.
    
    \bibliographystyle{apsrev}
    \bibliography{ref51019}
    
    \newpage
\end{document}